\documentclass[11pt,a4paper,twoside]{article}

\usepackage[latin1]{inputenc}
\usepackage[english]{babel}
\usepackage{amsmath,amssymb, enumerate,array,amsthm}
\newcommand{\Nesetril}[0]{Ne\-\v{s}e\-t\v{r}il}
\newcommand{\MC}[0]{\ensuremath{\textsc{MC}}}
\newcommand{\CSP}[0]{\ensuremath{\textsc{CSP}}}
\newcommand{\QCSP}[0]{\ensuremath{\textsc{QCSP}}}
\newcommand{\NP}[0]{\ensuremath{\mathsf{NP}}}
\newcommand{\coNP}[0]{\ensuremath{\mathsf{co\mbox{-}NP}}}

\newcommand{\Logspace}[0]{\ensuremath{\mathsf{Logspace}}}

\newcommand{\Ptime}[0]{\ensuremath{\mathsf{P}}}
\newcommand{\Pspace}[0]{\ensuremath{\mathsf{Pspace}}}

\newcommand{\NAE}[0]{\ensuremath{\mathsf{NAE}}}

\newcommand{\FO}[0]{\ensuremath{\mathbf{FO}}}
\newcommand{\tuple}[1]{\ensuremath{\mathbf{#1}}}

\newcommand{\notmodels}{\ \makebox[0.1cm][l]{\ensuremath{\models}}/ \ }

\theoremstyle{plain}
\newtheorem{theorem}{Theorem}%[section]

\newtheorem{lemma}[theorem]{Lemma}
\newtheorem{proposition}[theorem]{Proposition}

\theoremstyle{definition}

\theoremstyle{remark}

\title{ \Large \bf Dichotomies and Duality in First-order Model Checking Problems\footnote{This paper is available at http://arxiv.org/abs/cs.LO/0609022.}}
\author{ Barnaby Martin \\ {\small Department of Computer Science, University of Durham,} \\ {\small Science Labs, South Road, Durham DH1 3LE, U.K.} \\ {\tt b.d.martin@durham.ac.uk} }

\date{}

\begin{document}

% PTo have the good page numbers
\setcounter{page}{0}%%%....> I'll put the good number
% To get the pages at the top
\pagestyle{myheadings}

\maketitle

\begin{abstract}
We study the complexity of the model checking problem, for fixed model $A$, over certain fragments $\mathcal{L}$ of first-order logic. These are sometimes known as the expression complexities of $\mathcal{L}$. We obtain various complexity classification theorems for these logics $\mathcal{L}$ as each ranges over models $A$, in the spirit of the dichotomy conjecture for the Constraint Satisfaction Problem -- which itself may be seen as the model checking problem for existential conjunctive positive first-order logic.
\end{abstract}

\section{Introduction}
The \emph{model checking problem} over a logic $\mathcal{L}$ takes as input a structure $A$ and a sentence $\varphi$ of $\mathcal{L}$, and asks whether $A \models \varphi$. The problem can also be parameterised, either by the sentence $\varphi$, in which case the input is simply $A$, or by the model $A$, in which case the input is simply $\varphi$. Vardi has studied the complexity of this problem, principly for logics which subsume $\FO$, in \cite{VardiComplexity}. He describes the complexity of the unrestricted problem as the \emph{combined complexity}, and the complexity of the parameterisation by the sentence (respectively, model) as the \emph{data complexity} (respectively, \emph{expression complexity}). For the majority of his logics, the expression and combined complexities are comparable, and are one exponential higher than the data complexity. 

In this paper, we will be interested in taking certain fragments $\mathcal{L}$ of \FO, and studying the complexities of the parameterisation of the model checking problem by the model $A$, that is the expression complexities for certain $A$.
When $\mathcal{L}$ is the \emph{positive existential conjunctive} fragment of \FO, $\{ \wedge, \exists \}$-\FO, the model checking problem is equivalent to the much-studied \emph{constraint satisfaction problem} (\CSP). The parameterisation of this problem by the model $A$ is equivalent to what is sometimes described as the \emph{non-uniform} constraint satisfaction problem, $\CSP(A)$ \cite{KolaitisVardi}. It has been  conjectured \cite{Bulatov00:algebras,FederVardi} that the class of $\CSP$s exhibits \emph{dichotomy} -- that is, $\CSP(A)$ is always either in \Ptime\ or is \NP-complete, depending on the model $A$. This is tantamount to the condition that the expression complexity for $\{ \wedge, \exists \}$-\FO\ on $A$ is always either in \Ptime\ or is \NP-complete. While in general this conjecture remains open, it has been proved for certain classes of models $A$. Of particular interest to us is Hell and \Nesetril's dichotomy for undirected graphs $A$: in \cite{HellNesetril} it is proved that $\CSP(A)$ is in \Ptime, if $A$ has a self-loop or is bipartite, and is \NP-complete, if $A$ is any other undirected graph.

Owing to the natural duality between $\wedge, \exists$ and $\vee, \forall$, we  consider also various dual fragments. For example, the dual of  $\{ \wedge, \exists \}$-\FO\ is \emph{positive universal disjunctive} \FO, $\{ \vee, \forall \}$-\FO. It is straightforward to see that this class of expression complexities exhibits dichotomy between \Ptime\ and \coNP-complete if, and only if, the class of \CSP s exhibits dichotomy between \Ptime\ and \NP-complete.

This paper is organised as follows. Section 2 is devoted to preliminaries and Section 3 to fragments of \FO\ whose model checking problems are of low complexity. In Section 4, we consider those fragments that are related to \CSP s and their duals. In the case of positive existential conjunctive \FO, it makes little difference whether or not equality is allowed, that is the expression complexities for  $\{ \wedge, \exists \}$-\FO\ and $\{ \wedge, \exists, = \}$-\FO\ are equivalent. The same is not true of positive universal conjunctive \FO; while a classification of the expression complexities over $\{ \vee, \forall \}$-\FO\ is equivalent to the unproven \CSP\ dichotomy conjecture, we are able to give a full dichotomy for the expression complexities over $\{ \vee, \forall, = \}$-\FO. The reason for this is that the equality relation in the latter somehow simulates a disequality relation in the former. In Section 5, we consider fragments with a single quantifier, but both conjunction and disjunction; in all cases we are able to give dichotomies for the respective classes of expression complexities. Finally, in Section 6, we consider the scope for further work.

\section{Preliminaries}

In this paper, we consider only finite, non-empty relational structures. A signature $\sigma$ is a finite sequence of relation symbols $R_1,\ldots,R_j$, with respective arities $a_1,\ldots,a_j$. A $\sigma$-structure $A$ consists of a finite, non-empty set $|A|$ -- the universe or domain of $A$ -- together with some sets $R^A_1 \subseteq |A|^{a_1}, \ldots, R^A_j \subseteq |A|^{a_j}$. When the model $A$ is clear, we may drop the superscript, blurring the distinction between the \emph{actual} relation and the relation \emph{symbol}. We denote the cardinality of the domain of $A$ by $||A||$. We generally use $x,y,z$ to refer to elements of $A$, and $v_1,v_2,\ldots$ to refer to variables that range over those elements. A \emph{digraph} is any structure over the signature that contains a single, binary relation $E$. An \emph{undirected graph} is a digraph whose edge relation is symmetric; while in general we permit self-loops, we will often stipulate the futher restriction of antireflexivity.

Let $R^A$ be a $k$-ary relation on $A$. We say that $R^A$ is \emph{$x$-valid}, for some $x \in A$, if $x^k:=(x,\ldots,x) \in R^A$. If all $k$-tuples $\tuple{t} \in R^A$ are such that $\tuple{t}$ contains $k$ distinct elements of $A$, then we say $R^A$ is \emph{antireflexive}. On digraphs, the edge relation is antireflexive if, and only if, it is not $x$-valid for any $x$.

For a $\sigma$-structure $A$, we define its complement $\overline{A}$ as that structure having the same universe as $A$, but whose relations are the (set-theoretic) complements of the relations of $A$. That is, for each $R^A_i$, the relation $R^{\overline{A}}_i$ is defined by $\tuple{x} \in R^{\overline{A}}_i$ iff $\tuple{x} \notin R^A_i$.

For a $\sigma$-structure $A$ in which none of the relations $R^A_i$ is empty, we define the \emph{canonical relation} $R_A$ (note the subscript), of arity $a_1 + \ldots + a_j$ to be such that
\[
\begin{array}{l}
(v^1_1,\ldots,v^1_{a_1}, \ldots \ldots, v^j_1,\ldots,v^j_{a_j}) \in R_A \mbox{ iff} \\
(v^1_1,\ldots,v^1_{a_1}) \in R^A_1 \wedge \ldots \wedge (v^j_1,\ldots,v^j_{a_j}) \in R^A_j .\\
\end{array}
\]
For a $\sigma$-structure $A$ in which some of the relations are empty, define $R_A$ to be $R_{A'}$ where $A'$ is $A$ restricted to those relations that are non-empty (this will require restricting the signature).
If, and only if, all relations of $A$ are empty, then we set $R_A$ to be \O.

If $A$ and $B$ are $\sigma$-structures, then a \emph{homomorphism} from $A$ to $B$ is some function $h:|A|\rightarrow |B|$ s.t. for all relations $R_i$ and all $(t_1,\ldots,t_{a_i}) \in |A|^{a_i}$, if $(t_1,\ldots,t_{a_i}) \in R^A_i$ then $(h(t_1),\ldots,h(t_{a_i})) \in R^B_i$. We denote the existence of a homomorphism from $A$ to $B$ by $A \rightarrow B$. If we have both $A \rightarrow B$ and $B \rightarrow A$ then $A$ and $B$ are said to be \emph{homomorphically equivalent}. A \emph{retraction} of a structure $A$ is a homomorphism from $A$ to some induced substructure $B \subseteq A$; if such exists, $B$ is said to be a \emph{retract} of $A$. The \emph{core} of a structure $A$ is a minimal (w.r.t. size) retract of $A$ (we talk of \emph{the} core since it is readily proved that this is unique up to isomorphism). Let $K_n$ be the complete antireflexive digraph (i.e. clique) on $n$ vertices. We call an undirected graph \emph{bipartite} if its core is either $K_2$ or $K_1$.

We will be interested in fragments of first-order logic \FO\ both in the presence and absence of the natural binary equality relation $=$. Whenever we have $=$, it should be considered a bona fide extensional relation, i.e., for a structure $A$, it should occur in the canonical relation $R_A$. We consider \FO\ to be built over the alphabet $\Gamma_1 \cup \Gamma_0$, where $\Gamma_1 := \{ \neg, \wedge, \vee, \exists, \forall, = \}$ and $\Gamma_0 :=  \{ ( ,) ,R ,v, 0, 1 \}$,
in an inductive manner. For each $R_i$ in $\sigma$, and any natural numbers $j_1,\ldots,j_{a_i}$, $R_i(v_{j_1},\ldots,v_{j_{a_i}})$ is a \emph{formula} with \emph{free variables} $v_{j_1},\ldots,v_{j_{a_i}}$ (where each relation $R_i$ and variable $v_j$ is coded as $R \, bin(i)$ and $v \, bin(j)$, where $bin(i)$ and $bin(j)$ are the binary representations of $i$ and $j$, respectively). Likewise, for any $j_1,j_2$, $v_{j_1}=v_{j_2}$ is a formula with free variables $v_{j_1}, v_{j_2}$. If $\varphi$ and  $\psi$ are formulae, then $(\varphi \wedge \psi)$, $ (\varphi \vee \psi)$ and $(\neg \varphi)$ are also formulae, in each case having as free variables exactly those variables free in the constituent components. Finally, if the formula $\varphi$ contains the free variable $v_j$, then $(\exists v_j \varphi)$ and $(\forall v_j \varphi)$ are formulae, whose free variables are exactly those of $\varphi$ less $v_j$. A \emph{sentence} is a formula with no free variables. 

We will be interested in fragments of \FO\ that derive from restricting which of the symbols of $\Gamma_1 := \{ \neg, \wedge, \vee, \exists, \forall, = \}$ we permit.
In this paper we will concern ourselves with the non-trivial positive fragments involving exactly one quantifier. For $\Gamma \subseteq \Gamma_1$, we denote by $\Gamma$-\FO\ that fragment of \FO\ that is restricted to the symbols of $\Gamma \cup \Gamma_0$. We have $12$ cases to consider.

\[
\begin{array}{llll}
\mbox{Class I} & \mbox{Class II} & \mbox{Class III}  \\
\\
\{ \vee, \exists \} & \{ \wedge, \exists \} & \{ \wedge, \vee, \exists \}  \\
\{ \vee, \exists, = \}  & \{ \wedge, \exists, = \} & \{ \wedge, \vee, \exists, = \}  \\
\{ \wedge, \forall \} & \{ \vee, \forall \} & \{ \wedge, \vee, \forall \}  \\
\{ \wedge, \forall, = \} & \{ \vee, \forall, = \} & \{ \wedge, \vee, \forall, = \}  \\
\\
\end{array}
\]
For some $\Gamma \subseteq \Gamma_1$, and for some structure $A$, we define the \emph{model checking problem} $\Gamma$-$\MC(A)$ to have as input a sentence $\varphi$ of $\Gamma$-\FO, and as yes-instances those sentences such that $A \models \varphi$. The complexity of the model checking problem $\Gamma$-$\MC(A)$ may be termed the expression complexity for $\Gamma$-\FO\ on $A$, in line with the parlance of \cite{VardiComplexity}. The following is basic and may easily be verified.
\begin{lemma}
For each $\Gamma \subseteq \Gamma_1$, the recognition problem for well-formed sentences of $\Gamma$-$\FO$ is in \Logspace. Moreover, given any sentence $\varphi \in \Gamma$-$\FO$, we may compute in logarithmic space an equivalent sentence $\varphi'$ in prenex normal form.
\end{lemma}
In this paper, we will not be concerned with complexities beneath \Logspace. In light of the previous lemma, and w.l.o.g., we henceforth assume all inputs are in prenex normal form.

The following characterisations are hinted at in \cite{VardiComplexity}. Together with the dichotomy conjecture for \CSP, they provide much of the motivation for the present work.
\begin{proposition} \
\begin{itemize}
\item[$(i)$] In full generality, the class of problems $\{ \neg, \wedge, \vee, \exists, \forall, = \}$-$\MC(A)$, i.e. $\Gamma_1$-$\MC(A)$, exhibits dichotomy: if $||A||=1$ then the problem is in \Logspace, otherwise it is \Pspace-complete.
\item[$(ii)$] In full generality, the class of problems $\{ \neg, \wedge, \vee, \exists, \forall \}$-$\MC(A)$ exhibits dichotomy: if all relations of $A$ are either empty or contain all tuples (equivalently, $R_A$ is either empty or contains all tuples) then the problem is in \Logspace, otherwise it is \Pspace-complete.
\end{itemize}
\end{proposition}
\begin{proof}\hspace{-2mm}\footnote{
In \cite{VardiComplexity}, it is claimed that the \Pspace-hard cases of $(ii)$, which entail the \Pspace-hard cases of $(i)$, are proved in \cite{CM77}. We are unable to find such, and are in some doubt as to what would be an appropriate reference. Certainly $(i)$ qualifies as folklore, having been casually mentioned in \cite{KolaitisTalk}.
}
We sketch the proof for $(i)$; the proof for $(ii)$ is similar, if a little more involved. Note that \Pspace\ membership follows by a simple evaluation procedure inward through the quantifiers (see \cite{VardiComplexity}).

In the case where $||A||=1$, every relation is either empty or contains all tuples (one tuple), and the quantifiers $\exists$ and $\forall$ are semantically equivalent. Hence, the problem translates to the Boolean Sentence Value Problem (under the substitution of $0$ and $1$ for the empty and non-empty relations, respectively), known to be in \Logspace\ \cite{NLynch}.

When $||A|| \geq 2$, \Pspace-hardness may be proved using no extensional relation of $A$ other than $=$. The method involves a reduction from the \Pspace-complete Quantified Boolean Formula Problem (of \cite{StockmeyerPolynomialHierarchy}).
\end{proof}

\section{Logics of Class I}
We commence with the low-complexity logics of Class I. Let us consider the problem $\{ \vee, \exists \}$-$\MC(G)$, for some digraph $G$ of size $n$. An input for this problem will be of the form:
\[ \varphi \ := \ \exists \tuple{v} \ E(v_1,v'_1) \vee \ldots \vee E(v_m,v'_m) \]
where $v_1,v'_1, \ldots, v_m,v'_m$ are the not necessarily distinct variables that comprise $\tuple{v}$. Now, $G \models \varphi$ iff it contains an edge. The example demonstrates the triviality of the model checking problem on the fragment $\{ \vee, \exists \}$-\FO; the following proves it.
\begin{proposition}
Let $\Gamma$-\FO\ be any of the logics of Class I. For all structures $A$, the model checking problem $\Gamma$-$\MC(A)$ is in \Logspace.
\label{prop:vee-exists-equals}
\end{proposition}
\begin{proof}
For $\{ \vee, \exists \}$-$\MC(A)$ and $\{ \vee, \exists, = \}$-$\MC(A)$:
let $a$ be the maximum arity of the relations of $A$. Consider any prenex sentence for the model checking problem. Suppose $||A||=n$: we may cycle through each of the $n^a$ tuples in $|A|^a$ looking for a tuple that satisfies some disjunct (for relations of arity less than $a$, we consider prefix sub-tuples). If we find no such $a$-tuple then the input is a no-instance, otherwise it is a yes-instance. This requires space $a \log n$, and the result follows.

For $\{ \wedge, \forall \}$-$\MC(A)$ and $\{ \wedge, \forall, = \}$-$\MC(A)$, the proof is similar, except that we search for a tuple which falsifies some conjunct: if we find no such tuple, the input is a yes-instance, otherwise it is a no-instance.
\end{proof}

\section{Logics of Class II}

\subsection{$\{ \wedge, \exists \}$-\FO\ and $\{ \wedge, \exists, = \}$-\FO}
\label{sec:csp}

Owing to the rule of substitution, the logics $\{ \wedge, \exists \}$-\FO\ and $\{ \wedge, \exists, = \}$-\FO\ are very nearly identical. We have the trivial inclusion $\{ \wedge, \exists \}$-$\FO \subseteq \{ \wedge, \exists, = \}$-\FO. For the converse, consider any sentence of $\{ \wedge, \exists, = \}$-\FO\ that contains \emph{at least one} extensional relation that is not $=$. We may remove each instance of an equality $v_i=v_j$ and substitute all instances of $v_j$ with $v_i$ elsewhere in the sentence. Plainly, this sentence is equivalent to the original and is in $\{ \wedge, \exists \}$-\FO. We are left with the degenerate case of a sentence of $\{ \wedge, \exists, = \}$-\FO\ whose only relations are equalities. Such a sentence will be logically equivalent to $\exists v_1 \ v_1=v_1$, which is true on all models. It should be clear to the reader that the only structures on which $\{ \wedge, \exists \}$-\FO\ and $\{ \wedge, \exists, = \}$-\FO\ are not equivalent are those in which all relations are empty. It follows that for structures $A$ in which all relations are empty, while the problem $\{ \wedge, \exists \}$-$\MC(A)$ is genuinely trivial (has no yes-instances), the problem $\{ \wedge, \exists, = \}$-$\MC(A)$ is only very nearly trivial (sentences which contain only equalities form exactly the yes-instances). For the purposes of complexity analysis, we consider these logics equivalent.

The model checking problem $\{ \wedge, \exists \}$-$\MC(A)$ for inputs in prenex form is exactly the non-uniform constraint satisfaction problem $\CSP(A)$ (e.g. see \cite{OxfordQuantifiedConstraints}). We note that this problem is always in \NP: we may guess a satisfying assignment and verify in polynomial time. As we have mentioned, there is a conjectured dichotomy for $\CSP(A)$, namely that each instance is either in \Ptime\ or is \NP-complete \cite{Bulatov00:algebras,FederVardi}. This remains unproved. The following is a straightforward consequence of our definitions.
\begin{proposition}
The class of problems $\{ \wedge, \exists \}$-$\MC(A)$ exhibits dichotomy between those cases that are in \Ptime\ and those that are \NP-complete, if, and only if, the class of non-uniform constraint satisfaction problems $\CSP(A)$ exhibits the same dichotomy.
\end{proposition}

\subsection{$\{ \vee, \forall \}$-\FO}
\label{sec:cocsp}

This logic is dual to the logic $\{ \wedge, \exists \}$-\FO\ in the following sense. Consider a prenex sentence $\varphi$ of $\{ \vee, \forall \}$-\FO, where the variables among $\tuple{v_1},\ldots,\tuple{v_m}$ are exacly those of $\tuple{v}$:
\[ \varphi \ := \ \forall \tuple{v} \ R_{\alpha_1}(\tuple{v_1}) \vee \ldots \vee R_{\alpha_m}(\tuple{v_m}) \]
Now, $A \notmodels \varphi$ iff
\[
\begin{array}{lllll}
A \ \notmodels & & \forall \tuple{v} & R_{\alpha_1}(\tuple{v_1}) \vee \ldots \vee R_{\alpha_m}(\tuple{v_m}) & \mbox{ iff} \\
A \ \notmodels & \neg & \exists \tuple{v} & \neg[ R_{\alpha_1}(\tuple{v_1}) \vee \ldots \vee R_{\alpha_m}(\tuple{v_m})] & \mbox{ iff} \\
A \ \notmodels & \neg & \exists \tuple{v} & \neg R_{\alpha_1}(\tuple{v_1}) \wedge \ldots \wedge \neg R_{\alpha_m}(\tuple{v_m}) & \mbox{ iff} \\
A \ \models & & \exists \tuple{v} & \neg R_{\alpha_1}(\tuple{v_1}) \wedge \ldots \wedge \neg R_{\alpha_m}(\tuple{v_m}) & \mbox{ iff} \\
\overline{A} \ \models & & \exists \tuple{v} & R_{\alpha_1}(\tuple{v_1}) \wedge \ldots \wedge R_{\alpha_m}(\tuple{v_m}) & \mbox{ iff} \\
\end{array}
\]
$\overline{A} \models \varphi'$, where
\[ \varphi' \ := \ \exists \tuple{v} \ R_{\alpha_1}(\tuple{v_1}) \wedge \ldots \wedge R_{\alpha_m}(\tuple{v_m}) . \]
We can see that the problems $\{ \vee, \forall \}$-$\MC(A)$ and $\{ \wedge, \exists \}$-$\MC(\overline{A})$ are intimately related. Indeed, the complement of the problem $\{ \vee, \forall \}$-$\MC(A)$ is equivalent to the problem $\{ \wedge, \exists \}$-$\MC(\overline{A})$ under the reduction which swaps $\forall$ for $\exists$ and $\vee$ for $\wedge$. This reduction is extremely basic (certainly in \Logspace) and demonstrates that $\{ \vee, \forall \}$-$\MC(A)$ is always in \coNP. We also see that, if we choose some $A$ such that $\{ \wedge, \exists \}$-$\MC(\overline{A})$ is \NP-complete, then $\{ \vee, \forall \}$-$\MC(A)$ is \coNP-complete. The following is now elementary.
\begin{proposition}
The class of problems $\{ \vee, \forall \}$-$\MC(A)$ exhibits dichotomy (between \Ptime\ and \coNP-complete) if, and only if, the class of non-uniform constraint satisfaction problems $\CSP(A)$ exhibits dichotomy (between \Ptime\ and \NP-complete).
\end{proposition}

\subsection{$\{ \vee, \forall, = \}$-\FO}
\label{sec:cocspequ}

This logic is not dual to the logic $\{ \wedge, \exists, = \}$-\FO\ in the sense just described. Rather it is dual to the logic $\{ \wedge, \exists \}$-\FO\ when augmented with a \emph{disequality} relation. Since a disequality relation on a structure is tantamount to a graph clique, we are immediately led to the following.
\begin{proposition}
For structures $A$ such that $||A|| \geq 3$, the problem $\{ \vee, \forall, = \}$-$\MC(A)$ is \coNP-complete.
\label{prop:geq3.1}
\end{proposition}
\begin{proof}
For any $A$, membership of \coNP\ follows as in the previous section. Now let $||A||=n \geq 3$. We will prove that $\{ \vee, \forall, = \}$-$\MC(A)$ is \coNP-complete by reduction from the complement of the \NP-complete \emph{graph $n$-colourability} problem \cite{HellNesetril}, $\{ \wedge, \exists \}$-$\MC(K_n)$. Let an input for $\{ \wedge, \exists \}$-$\MC(K_n)$ be given, of the form:
\[ \varphi' \ := \ \exists \tuple{v} \ E(v_1,v'_1) \wedge \ldots \wedge E(v_m,v'_m) \]
where $v_1,v'_1, \ldots, v_m,v'_m$ are the not necessarily distinct variables that comprise $\tuple{v}$. Now, in a similar vein to the previous section, $K_n \notmodels \varphi'$ iff
\[
\begin{array}{llllll}
K_n & \notmodels & & \exists \tuple{v} &  E(v_1,v'_1) \wedge \ldots \wedge E(v_m,v'_m) & \mbox{ iff} \\
K_n & \ \models & & \forall \tuple{v} & \neg E(v_1,v'_1) \vee \ldots \vee \neg E(v_m,v'_m) & \mbox{ iff} \\
\overline{K_n} & \ \models & & \forall \tuple{v} & E(v_1,v'_1) \vee \ldots \vee E(v_m,v'_m) & \mbox{ iff} \\
K_n & \ \models & & \forall \tuple{v} & v_1=v'_1 \vee \ldots \vee v_m=v'_m & \mbox{ iff} \\
A & \ \models & & \forall \tuple{v} & v_1=v'_1 \vee \ldots \vee v_m=v'_m & \\
\end{array}
\]
which may be given as an input for the problem $\{ \vee, \forall, = \}$-$\MC(A)$.
\end{proof}
Thanks to Schaefer \cite{Schaefer} we can go further. For the further definitions required for the following, see the Appendix.
\begin{theorem}
\label{thm:cospequ}
In full generality, the class of problems $\{ \vee, \forall, = \}$-$\MC(A)$ exhibits dichotomy, between those cases that are in \Ptime\ and those that are \coNP-complete. Specifically:
\begin{itemize}
\item If $||A|| = 1$, then the problem $\{ \vee, \forall, = \}$-$\MC(A)$ is in $\Ptime$.
\item If $||A|| = 2$ then
  \begin{itemize}
  \item[] if $R_{\overline{A}}$ is $0$-valid, $1$-valid, horn, dual horn, bijunctive or affine, then $\{ \vee, \forall, = \}$-$\MC(A)$ is in $\Ptime$, otherwise
  \item[] $\{ \vee, \forall, = \}$-$\MC(A)$ is \coNP-complete.
  \end{itemize}
\item If $||A|| \geq 3$, then the problem $\{ \vee, \forall, = \}$-$\MC(A)$ is \coNP-complete.
\end{itemize}
\end{theorem}
\begin{proof}
For $||A||=1$ each relation $R^A_i$ is either empty or contains the single tuple $x^{a_i}$, where $x$ is the sole element of $A$. A sentence $\varphi:=$
\[ \forall \tuple{v} \ R_{\alpha_1}(\tuple{v_1}) \vee \ldots \vee R_{\alpha_m}(\tuple{v_m}) \]
(where the variables among $\tuple{v_1},\ldots,\tuple{v_m}$ are exactly those of $\tuple{v}$) may readily be evaluated on $A$ by forgetting the quantifiers and substituting for empty relations boolean false ($0$) and for non-empty relations boolean true ($1$). This leaves a boolean disjunction that is true iff it contains a disjunct $1$, i.e. iff $\varphi$ contains at least one non-empty relation. This is certainly verifiable in \Logspace. We specifically note that, if $\varphi$ contains an equality, i.e. a disjunct of the form $v=v'$, then $\varphi$ is certain to be true on $A$.

The case $||A|| \geq 3$ follows from the previous proposition. The case $||A||=2$ follows from our duality together with Schaefer's dichotomy theorem for generalised satisfiability \cite{Schaefer}.
\end{proof}

\section{Logics of Class III}

\subsection{$\{ \wedge, \vee, \exists \}$-\FO}

The logics $\{ \wedge, \vee, \exists \}$-\FO\ and $\{ \wedge, \vee, \exists, = \}$-\FO\ give rise to model checking problems whose internal structure is not dissimilar to those of $\{ \wedge, \exists \}$-\FO\ and $\{ \wedge, \exists, = \}$-\FO. For all four logics, the model checking problems are unique up to homomorphism equivalence.
\begin{lemma}
The following are equivalent.
\begin{itemize}
\item[$(i)$] The structures $A$ and $B$ are homomorphically equivalent.
\item[$(ii)$] The structures $A$ and $B$ have isomorphic cores.
\item[$(iii)$] The problems $\{ \wedge, \exists \}$-$\MC(A)$ and $\{ \wedge, \exists \}$-$\MC(B)$ coincide.
\item[$(iv)$] The problems $\{ \wedge, \exists, = \}$-$\MC(A)$ and $\{ \wedge, \exists, = \}$-$\MC(B)$ coincide.
\item[$(v)$] The problems $\{ \wedge, \vee, \exists \}$-$\MC(A)$ and $\{ \wedge, \vee, \exists \}$-$\MC(B)$ coincide.
\item[$(vi)$] The problems $\{ \wedge, \vee, \exists, = \}$-$\MC(A)$ and $\{ \wedge, \vee, \exists, = \}$-$\MC(B)$ coincide.
\end{itemize}
\end{lemma}
\begin{proof}
The equivalence of $(i)$ and $(ii)$ follows from the definitions. Each of the implications $(vi),(v),(iv) \rightarrow (i)$ follows from the well-documented $(iii) \rightarrow (i)$ \cite{FederVardi}.

For the remaining implications, it suffices to prove $(i) \rightarrow (vi)$. We can prove directly that $A \rightarrow B$ implies $\{ \wedge, \vee, \exists, = \}$-$\MC(A) \subseteq \{ \wedge, \vee, \exists, = \}$-$\MC(B)$ by appealing to the monotonicity of (the quantifier-free part of) $\{ \wedge, \vee, \exists, = \}$-$\FO$. The same applies with $B$ and $A$ swapped, and the result follows.
\end{proof}

We now turn our attention to $\{ \wedge, \vee, \exists \}$-\FO, returning to $\{ \wedge, \vee, \exists, = \}$-\FO\ in the next section.
\begin{proposition}
Let $G$ be an antireflexive digraph whose edge relation is non-empty. Then $\{ \wedge, \vee, \exists \}$-$\MC(G)$ is \NP-complete.
\label{prop:wedgeveeexists}
\end{proposition}
\begin{proof}
Membership of \NP\ remains elementary; we prove hardness. We may assume w.l.o.g. that $G$ is undirected (symmetric), since otherwise we may define the symmetric closure $E'$ of the edge relation $E$ via: $E'(u,v):=E(u,v)\vee E(v,u)$. More formally, $\{ \wedge, \vee, \exists \}$-$\MC(\mbox{sym-clos}(G))$ easily reduces to $\{ \wedge, \vee, \exists \}$-$\MC(G)$ under the reduction which substitutes instances $E^{\mbox{sym-clos}(G)}(u,v)$ in the former by $E^G(u,v) \vee E^G(v,u)$ in the latter.

Let $H$ be the core of $G$. Note that the \NP-hardness of $\{ \wedge, \exists \}$-$\MC(H)$ (a.k.a. $\CSP(H)$) immediately implies the \NP-hardness of both $\{ \wedge, \exists \}$-$\MC(G)$ and $\{ \wedge, \vee, \exists \}$-$\MC(G)$. Since $G$ is antireflexive and undirected, its core $H$ is either $K_1$ or $K_2$ or some non-bipartite $H'$.

The core $H$ can not be $K_1$, since then the edge relation of $G$ would have been empty.

If the core $H$ is a non-bipartite $H'$, then, by Hell and \Nesetril's theorem \cite{HellNesetril}, the problem $\{ \wedge, \exists \}$-$\MC(H')$ is \NP-complete, hence \NP-hardness of both $\{ \wedge, \exists \}$-$\MC(G)$ and \\ $\{ \wedge, \vee, \exists \}$-$\MC(G)$ follows.

It remains for us to consider the case where the core $H$ is $K_2$. By the previous lemma, it suffices for us to prove that $\{ \wedge, \vee, \exists \}$-$\MC(K_2)$ is \NP-hard. We define the ternary not-all-equal $\NAE_3$ relation on $K_2$ in $\{ \wedge, \vee, \exists \}$-$\FO$, whereupon we may appeal to the \NP-hardness of not-all-equal $3$-satisfiability (whose inputs may readily be expressed in $\{ \wedge, \vee, \exists \}$-$\FO$). We give $\NAE_3(u,v,w):=E(u,v)\vee E(v,w)\vee E(w,u)$.
\end{proof}

\begin{proposition}
Let $A$ be a structure whose canonical relation $R_A$ is $k$-ary for some $k \geq 2$. If $R_A$ is antireflexive, then $\{ \wedge, \vee, \exists \}$-$\MC(A)$ is \NP-complete.
\end{proposition}
\begin{proof}
Note that it follows from the defintion that $R_A$ is non-empty.
Consider the binary relation $E(v_1,v_2) := \exists v_3,\ldots,v_k R(v_1,\ldots,v_k)$. This relation specifies a non-empty, antireflexive digraph. The result follows from the previous proposition.
\end{proof}

\begin{proposition}
Let $A$ be a structure whose canonical relation $R_A$ is $k$-ary for some $k \geq 2$.
If $R_A$ is not $x$-valid, for all $x \in A$, then $\{ \wedge, \vee, \exists \}$-$\MC(G)$ is \NP-complete.
\end{proposition}
\begin{proof}
We take $R_A$ and build from it an antireflexive relation $R'$. Recall that $A$ is fixed, of size $||A||=n$, and consider its elements to be ordered $x_1,\ldots,x_n$. Take $R^{(0)}:=R_A$. From $R^{(m)}$, we build $R^{(m+1)}$ in the following manner. First, we list the tuples of $R^{(m)}$ lexicographically. We proceed through these tuples until we find one that has (at least one instance of) a repeated element. We now build $R^{(m+1)}$ by collapsing all the distinct repeated elements of that tuple to that distinct element. For example, if $R^{(m)}$ is of arity $5$, and the first tuple with the desired property is $(x_3,x_2,x_3,x_4,x_4)$, then $R^{(m+1)}(u,v,w):=R^{(m)}(u,v,u,w,w)$. Clearly this process terminates, i.e. a point $M$ is reached where $R^{(M)}=R^{(M+1)}$, and clearly $R^{(M)}$ is antireflexive. Furthermore, by non-$x$-validity of $R_A$, for all $x$, we know that $R^{(M)}$ has arity $k \geq 2$. We set $R':=R^{(M)}$. The result follows from the previous proposition.
\end{proof}

\begin{theorem}
The class of problems $\{ \wedge, \vee, \exists \}$-$\MC(A)$ exhibits dichotomy. Specifically,
if $R_A$ is either empty or $x$-valid, for some $x \in A$, then $\{ \wedge, \vee, \exists \}$-$\MC(A)$ is in \Logspace, otherwise
it is \NP-complete.
\end{theorem}

\begin{proof}
If $R_A$ is empty or $x$-valid, for some $x \in A$, then it follows that all relations $R^A_i$ are either empty or $x$-valid. An input for $\{ \wedge, \vee, \exists \}$-$\MC(A)$  readily translates to an input for the Boolean Sentence Value Problem, under the substitution of $0$ and $1$ for the empty and $x$-valid relations, respectively. The Boolean Sentence Value Problem is known to be in \Logspace\ \cite{NLynch}.

If $R_A$ is neither empty nor $x$-valid, for any $x \in A$, then the result follows from the previous proposition.
\end{proof}

\subsection{$\{ \wedge, \vee, \exists, = \}$-\FO}

This case is as the previous, via the same proof: although we may note that it is no longer possible for $R_A$ to be empty.
\begin{theorem}
The class of problems $\{ \wedge, \vee, \exists \}$-$\MC(A)$ exhibits dichotomy. Specifically,
if $R_A$ is $x$-valid, for some $x \in A$, then $\{ \wedge, \vee, \exists \}$-$\MC(A)$ is in \Logspace, otherwise
it is \NP-complete.
\end{theorem}

\subsection{$\{ \wedge, \vee, \forall \}$-\FO}

Note that the duality that we introduced in Section~\ref{sec:cocsp} works, via de Morgan's laws, perfectly well in the presence of both $\wedge$ and $\vee$. The case $\{ \wedge, \vee, \forall \}$-\FO\ is perfectly dual to $\{ \wedge, \vee, \exists \}$-\FO\ of the section before last. The following is straightforward.
\begin{theorem}
The class of problems $\{ \wedge, \vee, \forall \}$-$\MC(A)$ exhibits dichotomy. Specifically,
if $R_{\overline{A}}$ is either empty or $x$-valid, for some $x \in A$, then $\{ \wedge, \vee, \forall \}$-$\MC(A)$ is in \Logspace, otherwise
it is \coNP-complete. 
\end{theorem}

\subsection{$\{ \wedge, \vee, \forall, = \}$-\FO}

This logic is now dual to the logic $\{ \wedge, \vee, \exists \}$-\FO\ augmented with a disequality relation. 
\begin{theorem}
In full generality, the class of problems $\{ \wedge, \vee, \forall, = \}$-$\MC(A)$ exhibits trichotomy. Specifically:
\begin{itemize}
\item If $||A|| \geq 3$ then $\{ \wedge, \vee, \forall, = \}$-$\MC(A)$ is \coNP-complete.
\item If $||A \leq 2$ then: if $R_{\overline{A}}$ is either empty or $x$-valid, for some $x \in A$, then $\{ \wedge, \vee, \forall \}$-$\MC(A)$ is in \Logspace, otherwise
it is \coNP-complete.
\end{itemize}
\end{theorem}
\begin{proof}
The first part follows from Theorem~\ref{thm:cospequ} and the second part follows from the previous theorem.
\end{proof}

\section{Further Work}

In this paper we have not strayed into those fragments of \FO\ which contain both quantifiers. As has been mentioned, the model checking problem for \FO\ has been studied in \cite{VardiComplexity}, and dichotomies for $\{ \neg, \wedge, \vee, \exists, \forall, = \}$-$\MC(A)$ and $\{ \neg, \wedge, \vee, \exists, \forall \}$-$\MC(A)$ obtained.

Just as $\{ \wedge, \exists \}$-$\MC(A)$ is exactly the problem $\CSP(A)$, $\{ \wedge, \exists, \forall \}$-$\MC(A)$ is exactly the problem $\QCSP(A)$ -- the \emph{quantified constraint satisfaction problem} with template $A$. The classification problem for this class appears to be as difficult as that for the \CSP, and while it is known that complexities of \Ptime, \NP-complete\ and \Pspace-complete are attainable, little is known as to what may be inbetween. We anticipate that, with regard to model checking problems, the fragments $\{ \wedge, \exists, \forall, = \}$-\FO, $\{ \vee, \exists, \forall \}$-\FO\ and  $\{ \vee, \exists, \forall, = \}$-\FO\ sit in relation to $\{ \wedge, \exists, \forall \}$-\FO\ as the fragments $\{ \wedge, \exists, = \}$-\FO, $\{ \vee, \forall \}$-\FO\ and $\{ \vee, \forall, = \}$-\FO\ sit in relation to $\{ \wedge, \exists \}$-\FO.

This leaves the twin fragments $\{ \wedge, \vee, \exists, \forall \}$-\FO\ and $\{ \wedge, \vee, \exists, \forall, = \}$-\FO, whose model checking problems may demonstrate the richest variety of complexities. Certainly there are templates $A$ such that $\{ \wedge, \vee, \exists, \forall \}$-$\MC(A)$ attains each of the complexities \Ptime, \NP-complete, \coNP-complete and \Pspace-complete. However, a full classification resists.

\bibliographystyle{acm}
\bibliography{BarnyPaperLiterature}

\begin{thebibliography}{10}

\bibitem{OxfordQuantifiedConstraints}
{\sc Borner, F., Krokhin, A., Bulatov, A., and Jeavons, P.}
\newblock Quantified constraints and surjective polymorphisms.
\newblock Tech. Rep. PRG-RR-02-11, Oxford University, 2002.

\bibitem{Bulatov00:algebras}
{\sc Bulatov, A., Krokhin, A., and Jeavons, P.}
\newblock Constraint satisfaction problems and finite algebras.
\newblock In {\em Proceedings 27th International Colloquium on Automata,
  Languages and Programming, {ICALP'00}\/} (2000), vol.~1853 of {\em Lecture
  Notes in Computer Science}, Springer-Verlag, pp.~272--282.

\bibitem{CM77}
{\sc Chandra, A., and Merlin, P.}
\newblock Optimal implementation of conjunctive queries in relational
  databases.
\newblock In {\em {\em 9}th ACM Symposium on Theory of Computing\/} (1979),
  pp.~77--90.

\bibitem{FederVardi}
{\sc Feder, T., and Vardi, M.~Y.}
\newblock The computational structure of monotone monadic {SNP} and constraint
  satisfaction: a study through datalog and group theory.
\newblock {\em SIAM J. Comput. 28\/} (1999).

\bibitem{HellNesetril}
{\sc Hell, P., and Ne\v{s}et\v{r}il, J.}
\newblock On the complexity of {H}-coloring.
\newblock {\em J. Combin. Theory Ser. B 48\/} (1990).

\bibitem{KolaitisTalk}
{\sc Kolaitis, P.}
\newblock Csp and logic, 2006.
\newblock Tutorial at the International Workshop on Mathematics of Constraint
  Satisfaction, Oxford.

\bibitem{KolaitisVardi}
{\sc Kolaitis, P., and Vardi, M.}
\newblock Conjunctive-query containment and constraint satisfaction.
\newblock In {\em Proc. 17th ACM Symp. on Principles of Database Systems\/}
  (1998).

\bibitem{NLynch}
{\sc Lynch, N.}
\newblock Log space recognition and translation of parenthesis languages.
\newblock {\em Journal of the ACM 24\/} (1977), 583--590.

\bibitem{Schaefer}
{\sc Schaefer, T.}
\newblock The complexity of satisfiability problems.
\newblock In {\em STOC\/} (1978).

\bibitem{StockmeyerPolynomialHierarchy}
{\sc Stockmeyer, L.}
\newblock The polynomial-time hierarchy.
\newblock {\em Theoretical Computer Science 3\/} (1977).

\bibitem{VardiComplexity}
{\sc Vardi, M.}
\newblock Complexity of relational query languages.
\newblock In {\em 14th Symposium on Theory of Computation\/} (1982).

\end{thebibliography}

\section{Appendix: Schaefer's Boolean Relations}

A structure or relation is \emph{boolean} if its domain is of size $2$. Without loss of generality, we may assume that the elements of the domain are $0$ and $1$. We may refer to boolean relations by some propositional formula that expresses them, reading the propositional variables lexicographically, e.g. $[P \vee Q]$ expresses $\{ (0,1),(0,1),(1,1)\}$; $[P \neq Q]$ expresses $\{ (0,1),(1,0) \}$.
A boolean relation $R$, of arity $a$, is:
\begin{itemize}
\item[$(i)$] \emph{$0$-valid} iff it contains the tuple $(0^a)$.
\item[$(ii)$] \emph{$1$-valid} iff it contains the tuple $(1^a)$. 
\item[$(iii)$] \emph{horn} iff it may be expressed by a propositional formula in CNF where each clause has at most one positive literal.
\item[$(iv)$] \emph{dual horn} iff it may be expressed by a propositional formula in CNF where each clause has at most one negative literal.
\item[$(v)$] \emph{bijunctive} iff it may be expressed by a propositional formula in 2-CNF.
\item[$(vi)$] \emph{affine} iff it may be expressed by a propositional formula that is the conjunction of linear equations over $\mathbf{Z}_2$.
\end{itemize}
The following is Schaefer's dischotomy theorem for \emph{generalised satisfiability without constants}.
\begin{theorem}\cite{Schaefer}
Let $A$ be a boolean structure. Then $\CSP(A)$ (equivalently, $\{ \wedge, \exists \}$-$\MC(A)$) is in \Ptime\ if $R_A$ is in any of the six classes above, otherwise it is \NP-complete.
\end{theorem}

\end{document}